\documentclass[12pt]{article}
\usepackage{amsfonts}
\usepackage{epsfig}

\begin{document}

\title{\textbf{Option pricing with fractional volatility}}
\author{\textbf{R. Vilela Mendes}\thanks{%
Universidade T\'{e}cnica de Lisboa and Grupo de F{\'\i}sica Matem\'{a}tica,
e-mail: vilela@cii.fc.ul.pt}\quad and \textbf{Maria Jo\~{a}o Oliveira}%
\thanks{%
Universidade Aberta and Grupo de F{\'\i}sica Matem\'{a}tica, e-mail:
oliveira@cii.fc.ul.pt} \\
Complexo Interdisciplinar,\\
Av. Prof. Gama Pinto 2,1649-003 Lisboa, Portugal}
\date{}
\maketitle

\begin{abstract}
Based on empirical market data, a stochastic volatility model is proposed
with volatility driven by fractional noise. The model is used to obtain a
risk-neutrality option pricing formula and an option pricing equation.
\end{abstract}


\section{Introduction}

The Black-Scholes (B-S) \cite{Black} \cite{Merton} pricing formula for call
options is widely used by market agents. However, rather than as a
predictive formula, it is used as a means of quoting option prices in terms
of another parameter, the implied volatility. If $C_{BS}\left(
S_{t},K,T-t,\sigma _{t}\right) $ is the Black-Scholes price for an option at
time $t$, with strike price $K$, maturity time $T$ and underlying price $%
S_{t}$ then, the implied volatility $\sigma _{t}^{\mathnormal{imp}}$ is such
that 
\begin{equation}
C_{BS}\left( S_{t},K,T-t,\sigma _{t}^{\mathnormal{imp}}\right) =C_{t}\left(
K,T\right)  \label{1.01}
\end{equation}
$C_{t}\left( K,T\right) $ being the market price of a (liquid) option.

B-S is based on unrealistic assumptions about the market process: geometric
Brownian motion with constant non-stochastic volatility, continuous
adjustment of the portfolio and no transaction fees. Nevertheless B-S owes
its great popularity to its simplicity and to the fact that $C_{BS}$ is an
invertible map from $\left[ 0,\infty \right) $ to $\left[
0,S_{t}-Ke^{-r\left( T-t\right) }\right) $, $r$ being the risk-free rate.

The empirical shortcomings of B-S originated a vast amount of alternative
models \cite{Merton2} $-$ \cite{Bouchaud}, which attempt to explain the
deviations from B-S by introducing additional degrees of freedom. The
majority of these models have no compact closed-form solution and numerical
solutions most often do not reproduce correctly the data profiles \cite
{Bakshi} \cite{Tompkins}.

Given the success of the Black-Scholes formula as a compact invertible
parametrization, any alternative model that attempts to introduce a higher
degree of market realism should, at least, provide also a compact
closed-form recipe to parametrize the data.

In liquid markets the autocorrelation of price changes decays to negligible
quantities in a few minutes, consistent with the absence of long term
statistical arbitrage. Geometric Brownian motion models well this lack of
memory, although it does not reproduce the empirical leptokurtosis. On the
other hand, nonlinear functions of the returns exhibit significant positive
autocorrelation. For example, there is volatility clustering, with large
returns expected to be followed by large returns and small returns by small
returns (of either sign). This has the clear implication that, on stochastic
volatility models, long memory effects should be represented in the
volatility process.

Two-factor models with one persistent factor and one quickly mean-reverting
factor \cite{Alizadeh}, multicomponent GARCH models as the one proposed in 
\cite{Ding} (LM(q)-ARCH) or fractionally integrated processes \cite{Crato} 
\cite{Vilasuso} provide adequate fits to the empirical volatility data.
However, these models do not possess the analytical properties required to
construct an option pricing equation generalizing Black-Scholes.

Because the market process, as well as some functionals of their variables,
display approximate self-similar properties, mathematical simplicity
suggests to look for descriptions in terms of fractional Brownian motion. In
fact, if a nondegenerate process $X_{t}$ has finite variance, stationary
increments and is self-similar 
\begin{equation}
\mathnormal{Law}\left( X_{at}\right) =\mathnormal{Law}\left(
a^{H}X_{t}\right)  \label{1.02}
\end{equation}
then \cite{Embrechts} $0<H\leq 1$ and 
\begin{equation}
\mathnormal{Cov}\left( X_{s},X_{t}\right) =\frac{1}{2}\left( \left| s\right|
^{2H}+\left| t\right| ^{2H}-\left| s-t\right| ^{2H}\right) E\left(
X_{1}^{2}\right)  \label{1.03}
\end{equation}
The simplest process with these properties is a Gaussian process called
fractional Brownian motion. Fractional Brownian motion \cite{Mandelbrot1} 
\begin{equation}
\Bbb{E}\left[ B_{H}\left( t\right) \right] =0\qquad \Bbb{E}\left[
B_{H}\left( t\right) B_{H}\left( s\right) \right] =\frac{1}{2}\left\{ \left|
t\right| ^{2H}+\left| s\right| ^{2H}-\left| t-s\right| ^{2H}\right\}
\label{1.1}
\end{equation}
has for $H>\frac{1}{2}$ a long range dependence 
\begin{equation}
\sum_{n=1}^{\infty }\mathnormal{Cov}\left( B_{H}\left( 1\right) ,B_{H}\left(
n+1\right) -B_{H}\left( n\right) \right) =\infty  \label{1.2}
\end{equation}
and was suggested in the past as a tool for modeling in Finance \cite
{Mandelbrot2}. However, because it was pointed out \cite{Rogers} that
markets based on $B_{H}\left( t\right) $ could have arbitrage, fractional
Brownian motion was no longer considered, by many, as promising for
mathematical modeling in Finance. The arbitrage result in \cite{Rogers} is a
consequence of using pathwise integration. With a different definition \cite
{Duncan}, 
\begin{equation}
\int_{a}^{b}f\left( t,\omega \right) dB_{H}\left( t\right) =\lim_{\left|
\Delta \right| \rightarrow 0}\sum_{k}f\left( t_{k},\omega \right) \diamond
\left( B_{H}\left( t_{k+1}\right) -B_{H}\left( t_{k}\right) \right)
\label{1.3}
\end{equation}
where $\Delta :a=t_{0}<t_{1}<\cdots <t_{n}=b$ is a partition of the interval 
$[a,b]$, $\left| \Delta \right| =\max_{0\leq k\leq n-1}\left(
t_{k+1}-t_{k}\right) $ and $\diamond $ denotes the Wick product, the
integral has zero expectation value and the arbitrage result is no longer
true. This is, in fact, the most natural definition because it is the Wick
product that is associated to integrals of It\^{o} type, whereas the usual
product is natural for integrals of Stratonovich type.

An essentially equivalent approach constructs the stochastic integral
through the divergence operator and Malliavin calculus \cite{Nualart}. A
fully consistent stochastic calculus has therefore been developed for
fractional Brownian motion \cite{Ustunel} \cite{Duncan} \cite{Hu} \cite
{Oksendal1} \cite{Nualart} \cite{Alos}. We will draw on these results, in
particular on the fractional generalization of the It\^{o} formula, to
derive our stochastic differential market model and option pricing equation.

However, to simply postulate that the price process follows a stochastic law 
\begin{equation}
\begin{array}{lll}
dS_{t} & = & \mu S_{t}dt+\sigma S_{t}dB_{H}\left( t\right)
\end{array}
\label{1.4}
\end{equation}
with $B_{H}\left( t\right) $ ($H>1/2$) a persistent fractional Brownian
motion, as done in \cite{Shiryaev} \cite{Oksendal2}, although mathematically
pleasant, is not phenomenological correct in view of the fact that price
changes have a short memory. The simplest alternative would be to generalize
current stochastic volatility models by changing the Hurst coefficient of
the volatility process, namely 
\begin{equation}
\begin{array}{lll}
dS_{t} & = & \mu S_{t}dt+\sigma _{t}S_{t}dB\left( t\right) \\ 
d\sigma _{t} & = & \mu _{V}\left( \theta -\sigma _{t}\right) dt+\sigma
_{V}\sigma _{t}dB_{H}\left( t\right)
\end{array}
\label{1.5}
\end{equation}
The price would follow a geometrical Brownian process, whereas the
volatility would have a mean-reverting deterministic term and a stochastic
component modeled by a geometrical fractional Brownian motion of Hurst
coefficient $H>\frac{1}{2}$. This, on the one hand, would introduce a memory
component on the volatility (where the data says it should be) and, on the
other hand, as it is known \cite{Stein} \cite{LeBaron} stochastic volatility
changes the effective probability distribution function of $S_{t}$ (even for 
$H=\frac{1}{2}$), generating fat tails closer to the empirical data.

However, the stochastic volatility model of Eqs.(\ref{1.5}), as we will see
later, turns out not to be justified as well, because its scaling properties
are very different from those of the empirical data. For this reason we have
made a detailed analysis of the data, reported in Section 2, trying to infer
a model which is, at the same time, mathematically manageable and not
grossly disproved by the data. Our conclusion is the following coupled
stochastic system 
\begin{equation}
\begin{array}{lll}
dS_{t} & = & \mu S_{t}dt+\sigma _{t}S_{t}dB\left( t\right) \\ 
\log \sigma _{t} & = & \beta +\frac{k}{\delta }\left\{ B_{H}\left( t\right)
-B_{H}\left( t-\delta \right) \right\}
\end{array}
\label{1.6}
\end{equation}
It means that, in addition to a mean value, volatility is driven not by
fractional Brownian motion but by fractional noise. Notice that our
empirically based model is quite different from the usual stochastic
volatility models which assume the volatility to follow an arithmetic or
geometric Brownian process. $\delta $ is the observation scale of the
process. In the $\delta \rightarrow 0$ limit the driving process would be
the distribution-valued process $W_{H}$%
\begin{equation}
W_{H}=\lim_{\delta \rightarrow 0}\frac{1}{\delta }\left( B_{H}\left(
t\right) -B_{H}\left( t-\delta \right) \right)  \label{1.7}
\end{equation}
In (\ref{1.6}) the constant $k$ measures the strength of the volatility
randomness. Although phenomenologically grounded and mathematically well
specified, the stochastic system (\ref{1.6}) is still a limited model
because, in particular, the fact that the volatility is not correlated with
the price process excludes the modeling of leverage effects.

In Section 3, following a risk-neutrality approach \cite{Cox1} \cite{Hull},
we use the conditional probability distributions following from Eqs.(\ref
{1.6}) to derive an option price formula and compare it with Black-Scholes,
namely in terms of the equivalent implied volatility surfaces.

The risk-neutrality approach of Section 3 provides a fairly simple explicit
generalization of Black-Scholes, which we believe to be based in a more
realistic mathematical representation of the market process. Nevertheless it
relies on some approximations which may or may not be justified. Therefore,
for future reference, we derive in Section 4 an option pricing equation of
wider generality and obtain an integral representation of its solution.

\section{The volatility process. An empirical analysis}

Option pricing being our primary motivation, the relevant time scales and
temporal horizon is greater than one day. Therefore we will use daily data
from the New York Stock Exchange (NYSE), the aggregate index and individual
companies as well. To discount trend effects and approach asymptotic
stationarity of the processes, we have detrended and rescaled the data as
explained in Ref.\cite{Vilela1}.

The first objective is to check whether the assumption that the price
process follows a geometric Brownian process 
\begin{equation}
dS_{t}=S_{t}\left( \mu dt+\sigma dB_{t}\right)  \label{2.1}
\end{equation}
is an acceptable hypothesis for the daily data. For detrended data $\mu =0$
and if the stochastic part is a self-similar process it should satisfy 
\begin{equation}
E\left| \frac{S\left( t+\Delta \right) -S\left( t\right) }{S\left( t\right) }%
\right| \sim \Delta ^{H}  \label{2.2}
\end{equation}

\begin{figure}[htb]
\begin{center}
\psfig{figure=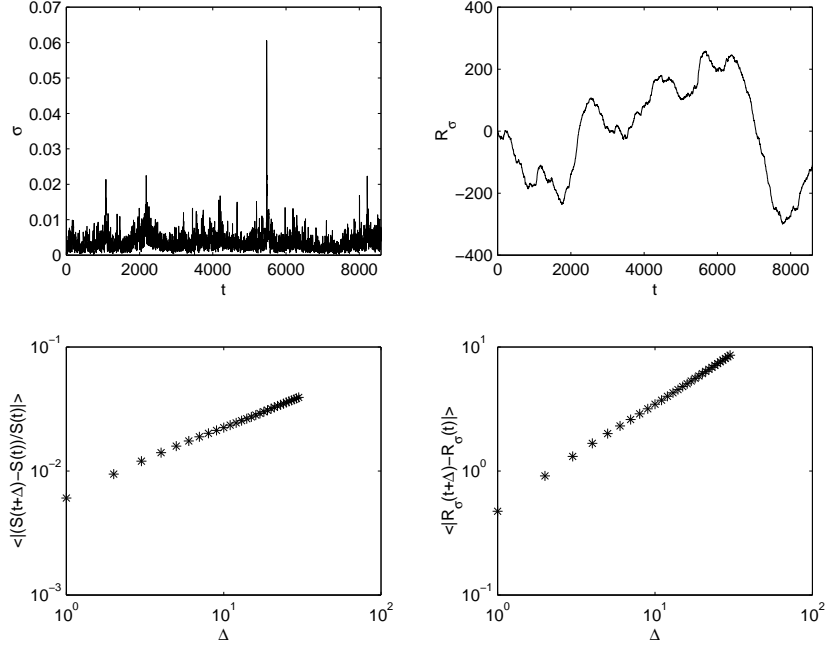,width=11truecm}
\end{center}
\caption{Self-similar properties of the price $S\left( t\right) $ and
integrated log-volatility $\beta t+R_{\sigma}\left( t\right) $ processes}
\end{figure}

In the lower left panel of Fig.1 one checks Eq.(\ref{2.2}) for the NYSE
index in the period 1966$-$2000 with $\Delta =1:30$ days. We obtain
approximate linearity in the log-log plot with $H=0.52$. There is a small
deviation from linearity for the first few days, but one should notice that
stochastic volatility effects have not yet been taken into account. So, the
geometric Brownian hypothesis for the daily price process is, at least, not
very unreasonable. It would imply that, in first approximation, the
probability distribution would be log-normal with volatility (variance)
given by 
\begin{equation}
\sigma _{t}^{2}=\frac{1}{\left| T_{0}-T_{1}\right| }\mathnormal{var}\left(
\log S_{t}\right)  \label{2.3}
\end{equation}
Eq.(\ref{2.3}) is used to extract the volatility process $\sigma _{t}$ from
the data and one checks whether a relation of the form 
\begin{equation}
E\left| \sigma \left( t+\Delta \right) -\sigma \left( t\right) \right| \sim
\Delta ^{H}  \label{2.4}
\end{equation}
or 
\begin{equation}
E\left| \frac{\sigma \left( t+\Delta \right) -\sigma \left( t\right) }{%
\sigma \left( t\right) }\right| \sim \Delta ^{H}  \label{2.5}
\end{equation}
hold for the volatility process. This would be the behavior implied by most
of the stochastic volatility models that have been proposed in the past. It
turns out that the data shows this to be a very bad hypothesis, meaning that
the volatility process itself is not self-similar. However, the integrated
log-volatility is well represented by a relation of the form 
\begin{equation}
\sum_{n=0}^{t/\delta }\log \sigma \left( n\delta \right) =\beta t+R_{\sigma
}\left( t\right)  \label{2.6}
\end{equation}
As shown in the lower right panel of Fig.1 the $R_{\sigma }\left( t\right) $
process has self-similar properties. We will identify it with a fractional
Brownian motion with Hurst coefficient $H\simeq 0.8$ (for the NYSE index), 
\begin{equation}
R_{\sigma }\left( t\right) =kB_{H}\left( t\right)  \label{2.6a}
\end{equation}
The same parametrization holds for the data of all individual companies that
we tested, with $H$ in the range $0.8-0.9$. This analysis bears some
resemblance to the detrended fluctuation analysis technique \cite{Stanley}.
Notice however that here, because we are using market data that has already
been detrended and rescaled, it is used merely to extract the mean
volatility.

Eq.(\ref{2.6}) provides a simple mathematical representation of the
volatility memory effects. In particular, it means that the volatility is
not driven by fractional Brownian motion but by fractional noise, 
\begin{equation}
\log \sigma _{t}=\beta +\frac{k}{\delta }\left( B_{H}\left( t\right)
-B_{H}\left( t-\delta \right) \right)  \label{2.7}
\end{equation}
$\delta $ being the observation time scale (one day, for daily data). This
provides a simple interpretation of the fact that empirical return
statistics depend on the observation time scale. In the $\delta \rightarrow
0 $ limit (continuous time resolution) the driving process would be the
distribution valued process (fractional white noise) 
\begin{equation}
W_{H}\left( t\right) =\lim_{\delta \rightarrow 0}\frac{1}{\delta }\left(
B_{H}\left( t\right) -B_{H}\left( t-\delta \right) \right)  \label{2.8}
\end{equation}
For the volatility (at resolution $\delta $) 
\begin{equation}
\sigma \left( t\right) =\theta e^{\frac{k}{\delta }\left\{ B_{H}\left(
t\right) -B_{H}\left( t-\delta \right) \right\} -\frac{1}{2}\left( \frac{k}{%
\delta }\right) ^{2}\delta ^{2H}}  \label{2.9}
\end{equation}
the term $-\frac{1}{2}\left( \frac{k}{\delta }\right) ^{2}\delta ^{2H}$
being included to insure that $E\left( \sigma \left( t\right) \right)
=\theta $.

Eqs. (\ref{2.1}) and (\ref{2.7}), or equivalently (\ref{2.1}) and (\ref{2.9}%
), define our stochastic volatility model.

This stochastic volatility model will be used in the next section to derive
an option pricing formula. However it is also useful to compute the
modifications implied by the stochastic volatility on the probability
distribution of the price returns. From (\ref{2.7}) one concludes that $\log
\sigma _{t}$ is a Gaussian process with mean $\beta $ and covariance 
\begin{equation}
\psi \left( s,u\right) =\frac{k^{2}}{2\delta ^{2}}\left\{ \left| s-u+\delta
\right| ^{2H}+\left| u-s+\delta \right| ^{2H}-2\left| s-u\right|
^{2H}\right\}  \label{2.10}
\end{equation}
This Gaussian process has non-trivial correlation for $H\neq \frac{1}{2}$.
At each fixed time $\log \sigma _{t}$ is a Gaussian random variable with
mean $\beta $ and variance $k^{2}\delta ^{2H-2}$. Then, 
\begin{equation}
p_{\delta }\left( \sigma \right) =\frac{1}{\sigma }p_{\delta }\left( \log
\sigma \right) =\frac{1}{\sqrt{2\pi }\sigma k\delta ^{H-1}}\exp \left\{ -%
\frac{\left( \log \sigma -\beta \right) ^{2}}{2k^{2}\delta ^{2H-2}}\right\}
\label{2.11}
\end{equation}
therefore

\begin{equation}
P_{\delta }\left( \log \frac{S_{T}}{S_{t}}\right) =\int_{0}^{\infty }d\sigma
p_{\delta }\left( \sigma \right) p_{\sigma }\left( \log \frac{S_{T}}{S_{t}}%
\right)  \label{2.12}
\end{equation}
with 
\begin{equation}
p_{\sigma }\left( \log \frac{S_{T}}{S_{t}}\right) =\frac{1}{\sqrt{2\pi
\sigma ^{2}\left( T-t\right) }}\exp \left\{ -\frac{\left( \log \left( \frac{%
S_{T}}{S_{t}}\right) -\left( \mu -\frac{\sigma ^{2}}{2}\right) \left(
T-t\right) \right) ^{2}}{2\sigma ^{2}\left( T-t\right) }\right\}
\label{2.13}
\end{equation}
One sees that the effective probability distribution of the returns depends
on the observation time scale $\delta $. This is a pleasant feature of this
stochastic volatility model, which contrasts, for example, with GARCH\
models which describe well the volatility at a given time resolution, but
fail to account for the $\delta $ dependence.

\section{Option pricing. Risk-neutrality approach}

Assuming risk neutrality \cite{Cox1}, the value $V\left( S_{t},\sigma
_{t},t\right) $ of an option must be the present value of the expected
terminal value discounted at the risk-free rate 
\begin{equation}
V\left( S_{t},\sigma _{t},t\right) =e^{-r\left( T-t\right) }\int V\left(
S_{T},\sigma _{T},T\right) p\left( S_{T}|S_{t},\sigma _{t}\right) dS_{T}
\label{3.1}
\end{equation}
$V\left( S_{T},\sigma _{T},T\right) =\max \left[ 0,S-K\right] $ and the
conditional probability for the terminal price depends on $S_{t}$ and $%
\sigma _{t}$. As in Hull and White \cite{Hull}, we make use of the relation
between conditional probabilities of related variables, namely 
\begin{equation}
p\left( S_{T}|S_{t},\sigma _{t}\right) =\int p\left( S_{T}|S_{t},\overline{%
\log \sigma }\right) p\left( \overline{\log \sigma }|\log \sigma _{t}\right)
d\left( \overline{\log \sigma }\right)  \label{3.2}
\end{equation}
$\overline{\log \sigma }$ being the random variable 
\begin{equation}
\overline{\log \sigma }=\frac{1}{T-t}\int_{t}^{T}\log \sigma _{s}ds
\label{3.3}
\end{equation}
Then Eq.(\ref{3.1}) becomes 
\begin{equation}
V\left( S_{t},\sigma _{t},t\right) =\int C\left( S_{t},e^{\overline{\log
\sigma }},t\right) p\left( \overline{\log \sigma }|\log \sigma _{t}\right)
d\left( \overline{\log \sigma }\right)  \label{3.4}
\end{equation}
\begin{equation}
C\left( S_{t},e^{\overline{\log \sigma }},t\right) =\int e^{-r\left(
T-t\right) }V\left( S_{T},\sigma _{T},T\right) p\left( S_{T}|S_{t},\overline{%
\log \sigma }\right) dS_{T}  \label{3.5}
\end{equation}
$C\left( S_{t},e^{\overline{\log \sigma }},t\right) $ being the
Black-Scholes price for an option with average volatility $e^{\overline{\log
\sigma }}$, which is known to be 
\begin{equation}
C\left( S_{t},\sigma ,t\right) =S_{t}\left( a+b\right) N\left( a,b\right)
-Ke^{-r\left( T-t\right) }\left( a-b\right) N\left( a,-b\right)  \label{3.6}
\end{equation}
with 
\begin{equation}
\begin{array}{lll}
a & = & \frac{1}{\sigma }\left( \frac{\log \frac{S}{K}}{\sqrt{T-t}}+r\sqrt{%
T-t}\right) \\ 
b & = & \frac{\sigma }{2}\sqrt{T-t}
\end{array}
\label{3.7}
\end{equation}
and 
\begin{equation}
N\left( a,b\right) =\frac{1}{\sqrt{2\pi }}\int_{-1}^{\infty }dye^{-\frac{%
y^{2}}{2}\left( a+b\right) ^{2}}  \label{3.8}
\end{equation}

To compute the conditional probability $p\left( \overline{\log \sigma }|\log
\sigma _{t}\right) $ we recall that from our stochastic volatility model 
\begin{equation}
\overline{\log \sigma }=\log \sigma _{t}+\frac{1}{T-t}\int_{t}^{T}\frac{k}{%
\delta }ds\int_{t}^{s}\left( dB_{H}\left( \tau \right) -dB_{H}\left( \tau
-\delta \right) \right)  \label{3.9}
\end{equation}
Notice that, because we want to compute the conditional probability of $%
\overline{\log \sigma }$ given $\log \sigma _{t}$ at time $t$, $\sigma _{t}$
in Eq.(\ref{3.9}) is not a process but simply the value of the argument in
the $V\left( S_{t},\sigma _{t},t\right) $ function.

As a $t-$dependence process the double integral in (\ref{3.9}) is a centered
Gaussian process. Therefore, given $\log \sigma _{t}$ at time $t$, $%
\overline{\log \sigma }$ is a Gaussian variable with conditional mean and
variance 
\begin{equation}
E\left\{ \overline{\log \sigma }|\log \sigma _{t}\right\} =\log \sigma _{t}
\label{3.10}
\end{equation}
\begin{equation}
\alpha ^{2}=E\left\{ \left( \overline{\log \sigma }-\log \sigma _{t}\right)
^{2}\right\} =\frac{k^{2}}{\delta ^{2}\left( T-t\right) }\left\{ \frac{1}{%
2\left( T-t\right) }I_{1}+I_{2}\right\} +k^{2}\delta ^{2H-2}  \label{3.11}
\end{equation}
with 
\begin{equation}
I_{1}=\frac{2}{\left( 2H+1\right) \left( 2H+2\right) }\left\{ \left(
T-t+\delta \right) ^{2H+2}+\left( T-t-\delta \right) ^{2H+2}-2\left(
T-t\right) ^{2H+2}-2\delta ^{2H+2}\right\}  \label{3.12}
\end{equation}
\[
I_{2}=\frac{1}{2H+1}\left\{ 2\left( T-t\right) ^{2H+1}-\left( T-t+\delta
\right) ^{2H+1}-\left( T-t-\delta \right) ^{2H+1}\right\} 
\]
Because in general $\delta \ll \left( T-t\right) $ one obtains 
\begin{equation}
\alpha ^{2}\simeq \frac{k^{2}}{\delta ^{2-2H}}\left( 1-\left( 2H-1\right)
\left( \frac{\delta }{T-t}\right) ^{2-2H}\right)  \label{3.13}
\end{equation}
Finally 
\begin{equation}
p\left( \overline{\log \sigma }|\log \sigma _{t}\right) =\frac{1}{\sqrt{2\pi 
}\alpha }\exp \left\{ \frac{-\left( \overline{\log \sigma }-\log \sigma
_{t}\right) ^{2}}{2\alpha ^{2}}\right\}  \label{3.14}
\end{equation}
and from (\ref{3.4}) 
\begin{equation}
V\left( S_{t},\sigma _{t},t\right) =\int_{-\infty }^{\infty }d\xi C\left(
S_{t},e^{\xi },t\right) p\left( \xi |\log \sigma _{t}\right)  \label{3.15}
\end{equation}
one obtains 
\begin{equation}
V\left( S_{t},\sigma _{t},t\right) =S_{t}\left[ aM\left( \alpha ,a,b\right)
+bM\left( \alpha ,b,a\right) \right] -Ke^{-r\left( T-t\right) }\left[
aM\left( \alpha ,a,-b\right) -bM\left( \alpha ,-b,a\right) \right]
\label{3.16}
\end{equation}
\begin{eqnarray}
M\left( \alpha ,a,b\right) &=&\frac{1}{2\pi \alpha }\int_{-1}^{\infty
}dy\int_{0}^{\infty }dxe^{-\frac{\log ^{2}x}{2\alpha ^{2}}}e^{-\frac{y^{2}}{2%
}\left( ax+\frac{b}{x}\right) ^{2}}  \label{3.17} \\
&=&\frac{1}{4\alpha }\sqrt{\frac{2}{\pi }}\int_{0}^{\infty }dx\frac{e^{-%
\frac{\log ^{2}x}{2\alpha ^{2}}}}{ax+\frac{b}{x}}\textnormal{erf}\mathnormal{c}%
\left( -\frac{ax}{\sqrt{2}}-\frac{b}{\sqrt{2}x}\right)  \nonumber
\end{eqnarray}
as the new option price formula ($\textnormal{erf}$c being the complementary error
function and $a$ and $b$ being defined in Eq.(\ref{3.7})), with $\sigma $
replaced by $\sigma _{t}$.

\begin{figure}[htb]
\begin{center}
\psfig{figure=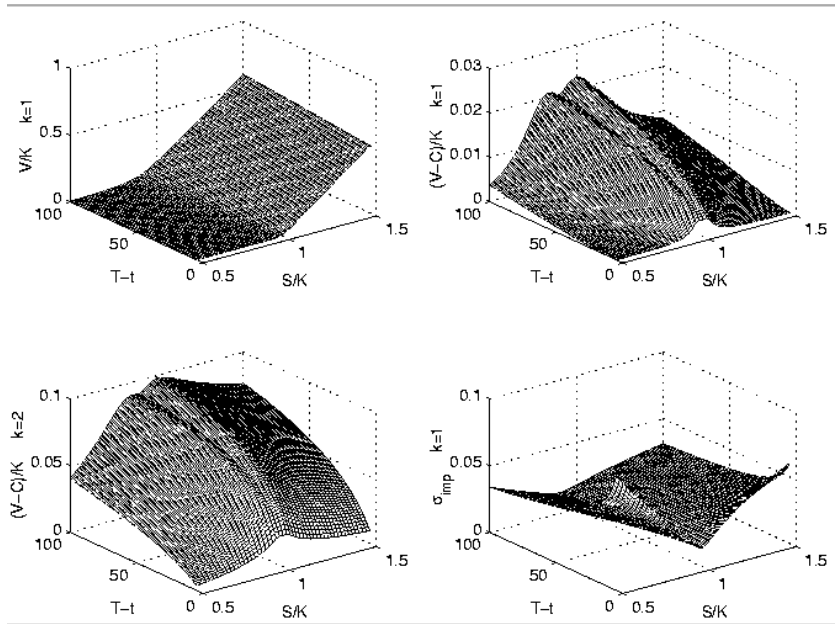,width=11truecm}
\end{center}
\caption{Option price and equivalent implied volatility in the risk-neutral
approach to the stochastic volatility model}
\end{figure}

In Fig.2 we plot the option value surface for $V\left( S_{t},\sigma
_{t},t\right) $ in the range $T-t\in [5,100]$ and $S/K\in [0.5,1.5]$ as well
as the difference $\left( V\left( S_{t},\sigma _{t},t\right) -C\left(
S_{t},\sigma _{t},t\right) \right) /K$ for $k=1$ and $k=2$. The other
parameters are fixed at $\sigma =0.01,r=0.001,\delta =1,H=0.8$. The lower
right panel shows the implied volatility surface corresponding to $V\left(
S_{t},\sigma _{t},t\right) $ for $k=1$. Notice that computationally it is
perhaps more convenient to use Eq.(\ref{3.15}) (and an optimized
Black-Scholes routine) than the more elegant Eq.(\ref{3.16}).

\section{An option pricing equation}

Whenever the actual drift of a financial time series can be replaced by the
risk-free rate we are in a risk-neutral situation. However, in stochastic
volatility models this may not be a good assumption. In particular, an
accurate option pricing formula should take into account the market price of
volatility risk. Here we derive an option pricing equation which does not
assume risk-neutrality.

Because volatility is not a tradable security, a pure arbitrage argument
cannot determine completely the fair price of an option. On the other hand,
because of the fractional nature of the volatility process, volatility
follows a stochastic process different from the one of the underlying
security. Therefore, we cannot apply the reasoning \cite{Lyuu} that leads to
uniform coefficients of the form $\left( \mu _{i}-\lambda _{i}\sigma
_{i}\right) $ in the first derivative terms of the option pricing equation%
\footnote{$\mu _{i}$, $\sigma _{i}$ and $\lambda _{i}$ would be the drift,
volatility and market price of risk for each process}. Hence, a first
principles derivation, with clearly specified assumptions is required.

As in Black-Scholes, we form a portfolio 
\begin{equation}
\Pi \left( t\right) =V\left( S,\sigma ,t\right) -\Delta \left( S,\sigma
,t\right) S_{t}  \label{4.1}
\end{equation}
From (\ref{2.1}) and (\ref{2.7}) and choosing $\Delta \left( S,\sigma
,t\right) =\frac{\partial V}{\partial S}$ we obtain 
\begin{eqnarray}
d\Pi \left( t\right) &=&\left\{ \frac{\partial V}{\partial t}+\frac{1}{2}%
\frac{\partial ^{2}V}{\partial S^{2}}\sigma ^{2}S^{2}\right\} dt+\sigma 
\frac{\partial V}{\partial \sigma }\frac{k}{\delta }\left( dB_{H}\left(
t\right) -dB_{H}\left( t-\delta \right) \right)  \label{4.2} \\
&&+\left( \sigma ^{2}\frac{\partial ^{2}V}{\partial \sigma ^{2}}+\sigma 
\frac{\partial V}{\partial \sigma }\right) H\frac{k^{2}}{\delta ^{2}}\delta
^{2H-1}dt  \nonumber
\end{eqnarray}
Consistent with our stochastic volatility model (\ref{1.6}), we have assumed
the volatility to be uncorrelated with $S$ and Eq.(\ref{4.2}) follows from
the application of the fractional It\^{o} formula \cite{Duncan} \cite
{Biagini}. Namely, if $X_{t}=\left( X_{t}^{(1)},X_{t}^{(2)},\cdots
,X_{t}^{(n)}\right) $ with $dX_{t}^{(i)}=c_{i}\left( t,\omega \right)
dB_{H}^{(i)}\left( t\right) $ , then 
\begin{equation}
df\left( t,X_{t}\right) =\frac{\partial f}{\partial t}dt+\sum_{i}\frac{%
\partial f}{\partial X^{\left( i\right) }}dX_{t}^{(i)}+\sum_{i}\frac{%
\partial ^{2}f}{\partial X^{\left( i\right) 2}}c_{i}\left( t,\omega \right)
D_{i,t}^{\phi }\left( X_{t}\right)  \label{4.3}
\end{equation}
$D_{i,t}^{\phi }\left( X_{t}\right) $ being the $\phi -$Malliavin derivative
corresponding to the $X_{t}^{(i)}-$process, defined by 
\begin{eqnarray}
D_{i,f}^{\phi }X_{t}\left( \omega _{i}\right) &=&\lim_{\varepsilon
\rightarrow 0}\frac{1}{\varepsilon }\left\{ X\left( \omega _{i}+\varepsilon
\int_{0}^{\bullet }ds\int_{0}^{\infty }\phi \left( s,u\right) f\left(
u\right) du\right) -X\left( \omega _{i}\right) \right\}  \label{4.4} \\
&=&\int_{0}^{\infty }D_{i,u}^{\phi }\left( X_{t}\right) f\left( u\right) du 
\nonumber
\end{eqnarray}
$\phi \left( s,u\right) $ being the kernel 
\begin{equation}
\phi \left( s,u\right) =H_{i}\left( 2H_{i}-1\right) \left| s-u\right|
^{2H_{i}-2}  \label{2.10a}
\end{equation}

In (\ref{4.2}) we are still left with the stochastic term $\sigma \frac{%
\partial V}{\partial \sigma }\frac{k}{\delta }\left( dB_{H}\left( t\right)
-dB_{H}\left( t-\delta \right) \right) $ and, because volatility is not a
tradable security this term cannot be eliminated by a portfolio choice.
Instead we may assume as reasonable to equate the deterministic term in $%
d\Pi \left( t\right) $ to 
\begin{equation}
\left( r\Pi \left( t\right) +\nu \frac{k}{\delta }\sigma \frac{\partial V}{%
\partial \sigma }\right) dt  \label{4.7}
\end{equation}
where $r$ is the risk-free return and, with $\nu >0$, the second term is a
measure of the market price of volatility risk (bigger risk, bigger return).
We end up with 
\begin{equation}
\frac{\partial V}{\partial t}+rS\frac{\partial V}{\partial S}+\frac{\sigma
^{2}S^{2}}{2}\frac{\partial ^{2}V}{\partial S^{2}}+\frac{k}{\delta }\left(
kH\delta ^{2H-2}-\nu \right) \sigma \frac{\partial V}{\partial \sigma }%
+Hk^{2}\delta ^{2H-3}\sigma ^{2}\frac{\partial ^{2}V}{\partial \sigma ^{2}}%
=rV  \label{4.8}
\end{equation}
as the general form of the option pricing equation consistent with the
stochastic volatility model in (\ref{1.6}).

We now obtain an integral representation for the solution of this equation.
With the change of variable 
\begin{equation}
x=\log \frac{S}{K}  \label{4.9}
\end{equation}
and passing to the two-dimensional Fourier transform 
\begin{equation}
V\left( t,x,\sigma \right) =\int \int d\phi d\rho F\left( \phi ,\rho ,\sigma
\right) e^{i\left( \phi t+\rho x\right) }  \label{4.10}
\end{equation}
we obtain 
\begin{equation}
Hk^{2}\delta ^{2H-3}\sigma ^{2}\frac{\partial ^{2}F}{\partial \sigma ^{2}}+%
\frac{k}{\delta }\left( kH\delta ^{2H-2}-\nu \right) \sigma \frac{\partial F%
}{\partial \sigma }+\left( i\left( \phi +\rho r-\frac{\sigma ^{2}\rho }{2}%
\right) -\frac{\sigma ^{2}\rho ^{2}}{2}-r\right) F=0  \label{4.11}
\end{equation}

Now, defining new constants 
\begin{equation}
\begin{array}{lll}
\chi \left( \rho \right) & = & \frac{\nu }{2Hk\delta ^{2H-2}} \\ 
\xi ^{2}\left( \rho ,\phi \right) & = & \chi ^{2}\left( \rho \right) -\frac{%
r-i\left( \phi +\rho r\right) }{Hk^{2}\delta ^{2H-3}} \\ 
\zeta ^{2}\left( \rho \right) & = & -\frac{i\rho +\rho ^{2}}{2Hk^{2}\delta
^{2H-3}}
\end{array}
\label{4.12}
\end{equation}
and making the replacement 
\begin{equation}
F\left( \sigma \right) =\sigma ^{\chi }Z_{\xi }\left( \zeta \sigma \right)
\label{4.13}
\end{equation}
Eq.(\ref{4.11}) reduces to a standard Bessel equation. Therefore the
solution of (\ref{4.8}) is 
\begin{equation}
V\left( t,x,\sigma \right) =\int \int d\rho d\phi e^{i\left( \phi t+\rho
x\right) }\sigma ^{\chi \left( \rho \right) }Z_{\xi (\rho ,\phi )}\left(
\zeta \left( \rho \right) \sigma \right)  \label{4.14}
\end{equation}
$Z_{\xi }\left( \zeta \sigma \right) $ being a Bessel function. The Bessel
function will be a linear combination 
\[
Z_{\xi }\left( \zeta \sigma \right) =c_{1}J_{\xi }\left( \zeta \sigma
\right) +c_{2}N_{\xi }\left( \zeta \sigma \right) 
\]
of a Bessel function of first kind and a Neumann function, with coefficients 
$c_{1}$ and $c_{2}$ to be fixed by the boundary conditions, which for call
options is 
\[
V\left( T,x,\sigma \right) =\max \left( x,0\right) 
\]

Eq.(\ref{4.14}) is an exact solution of the option pricing equation. How
useful it might be will depend on the development of adequate computational
algorithms for its numerical evaluation.

\end{document}